\documentclass[useAMS,usenatbib]{mn2e}
\usepackage{graphicx,amssym}
\newcommand{\bc}{\begin{center}}
\newcommand{\ec}{\end{center}}
\usepackage{color}
\newcommand{\comment}[1]{}
\title[The scatter in the stellar mass--halo mass relation]
      {On the scatter in the relation between stellar mass and halo mass: random or halo
      formation time dependent?}

\author[L.Wang, G.De Lucia, S.M.Weinmann]
       { Lan Wang$^{1}$$\thanks{Email: wanglan@bao.ac.cn}$,
	 Gabriella De Lucia$^{2}$,
        Simone M. Weinmann$^{3}$
	\\
        $^1$Partner Group of the Max Planck Institute for Astrophysics, National Astronomical Observatories, \\
        Chinese Academy of Sciences, 
  20A Datun Road, Chaoyang District, Beijing, China\\
	$^2$INAF - Astronomical Observatory of Trieste, via G.B. Tiepolo 11, I-34143 Trieste, Italy\\
	$^3$Leiden Observatory, Leiden University, P.O. Box 9513, 2300 RA Leiden, 
  The Netherlands   }
\begin{document}

\date{Accepted 2012 ???? ??. 
      Received 2012 ???? ??; 
      in original form 2012 ???? ??}

\pagerange{\pageref{firstpage}--\pageref{lastpage}} 
\pubyear{2012}

\maketitle

\label{firstpage}

\begin{abstract}
The empirical HOD
model of Wang et al. 2006 fits, by construction, both the stellar mass function 
and correlation function of galaxies in the local Universe. In contrast, the
semi-analytical models of De Lucia \& Blazoit 2007 (DLB07) and Guo et al. 2011 
(Guo11), built on the same dark matter halo merger trees than the empirical model, 
still have difficulties in reproducing these observational data simultaneously. 
We compare the relations between the stellar mass of galaxies and their host halo 
mass in the three models, and find that they are different. When the relations 
are rescaled to have the same median values and the same scatter as in Wang et al., 
the rescaled DLB07 model can fit both the measured galaxy stellar mass function 
and the correlation function measured in different galaxy stellar mass bins.
In contrast, the rescaled Guo11 model still over-predicts the clustering of low-mass 
galaxies. This indicates that the detail of how galaxies populate the scatter in the 
stellar mass -- halo mass relation does play an important role in determining the 
correlation functions of galaxies. While the stellar mass of galaxies in the Wang 
et al. model depends only on halo mass and is randomly distributed within the scatter,
galaxy stellar mass depends also on the halo formation time in semi-analytical models. 
At fixed value of infall mass, galaxies that lie above the median stellar mass -- 
halo mass relation reside in haloes that formed earlier, while galaxies that 
lie below the median relation reside in haloes that formed later. This effect is much 
stronger in Guo11 than in DLB07, which explains the over-clustering of low mass galaxies 
in Guo11. Assembly bias in Guo11 model might be overly strong. Nevertheless,
in case that a significant assembly bias indeed exists in the real Universe, 
one needs to use caution when applying
current HOD and abundance matching models that employ the assumption of 
random scatter in the relation between stellar and halo mass.

\end{abstract}

\begin{keywords}
   galaxies:haloes  -- galaxies: formation -- cosmology: large-scale structure of Universe
\end{keywords}

%%%%%%%%%%%%%%%%%%%%%%%%%%%%%%%%%%%%%%%%%
\section{Introduction}
\label{sec:intro}

In the currently favoured scenario for structure formation, galaxies are
believed to form by gas condensation within the potential wells of dark matter
haloes that form and evolve in a hierarchical bottom-up fashion: small haloes
form first and later merge to form more massive systems. Different methods have
been developed to link the physical properties of galaxies (such as their
stellar mass and/or luminosity) to the properties of their host haloes. 
These methods include measurements of halo masses of central galaxies through
e.g. galaxy-galaxy weak lensing \citep{mandelbaum2006} and satellite
kinematics \citep{more2009, more2009b, more2011}, and other methods that take advangage
of results from numerical simulations. Among these,
the traditional Halo Occupation Distribution (HOD) models have the following
ingredients: (i) the probability distribution relating the mass of a dark
matter halo to the number of galaxies that form within that halo, and (ii)
the spatial distribution of galaxies within their parent halo \citep{benson2000,
peacock2000, seljak2000, berlind2002, berlind2003}.  

The most recent renditions of this approach take advantage of high resolution
cosmological simulations to link the physical properties of galaxies to the
dynamical properties of dark matter substructures. As subhaloes fall into a
larger structure, they are subject to stripping and tidal disruption that
efficiently reduce their mass. Therefore, it is natural to assume that the
mass/luminosity of galaxies that reside within these substructures is
correlated with the subhalo mass at the time of `infall' ($M_{\rm infall}$),
i.e. at the time when the galaxy is, for the last time, a central galaxy of its own
halo \citep{vale2005, conroy2005, wang2006}. The most commonly used
observations to constrain the connection between galaxy properties and dark
matter haloes are the galaxy stellar mass/luminosity function, and the galaxy
correlation function \citep{yang2003, zehavi2005}.  The ``abundance matching''
methods use only the galaxy stellar mass function (SMF) as a constraint, and derive
the $M_{\rm star}$ -- $M_{\rm infall}$ relation assuming a monotonic
relationship between galaxy mass and halo mass \citep{moster2010, guo2010}. 
  
For the $M_{\rm star}$ -- $M_{\rm infall}$ relation, some models assume
no scatter in the relation \citep{guo2010}, while other models account for a
random scatter around the median relation \citep{wang2006, moster2010}. In
fact, one would naturally expect that, as detected and constrained in different
studies \citep{mandelbaum2006,more2011, skibba2011}, some scatter exists around the median
relation, due to the scatter in the formation and growth histories of dark matter
haloes \citep{zhao2003, liyun2007,zhao2009}, stochastic processes at play in
galaxy formation and evolution \citep{valle2005, kauffmann2006}, and
environmental physical processes \citep{goto2003, tanvuia2003, jaffe2011}.
Therefore, one could expect the scatter to be related to the physical properties 
of the parent dark matter haloes.
When modelling the scatter in the $M_{\rm star}$ -- $M_{\rm infall}$ relation,
however, all authors have so far assumed a Gaussian distribution in logarithmic
stellar mass \citep{wang2006, moster2010}: for a given halo mass, galaxy
stellar masses are equally and randomly assigned within the scatter,
independently of other halo properties.

While the HOD approach assumes that the galaxy content of a halo depends only
on its mass, recent studies have demonstrated that the clustering of dark
matter haloes depends on their formation time (usually defined as the time when
half of the final mass of the halo is first assembled in a single object). This
`assembly bias' was first pointed out by \citet{gao2005} who used a large
high-resolution simulation of the concordance $\Lambda$ cold dark matter
cosmogony to demonstrate that haloes less massive than about
$10^{13}\,h^{-1}\,{\rm M}_{\odot}$ that assembled at high redshift are
significantly more clustered than those of the same mass that assembled more
recently. Subsequent studies by \citet{zhu2006} and \citet{croton2007} studied
the dependence of galaxy properties on halo formation time using different
galaxy formation models. In particular, they found a dependence on halo formation 
time of galaxy clustering,
galaxy occupation number, and luminosity and stellar mass of central
galaxies. In addition, the stellar mass of satellite galaxies
also appears to depend on the FOF group mass at \emph{z}=0 \citep{neistein2011b}.

An alternative method to study galaxy formation and evolution is provided by
semi-analytic models (SAMs) \citep{white1991}. Unlike HODs that provide an
empirical/statistical relation between galaxy properties and host halo mass,
SAMs attempt to describe the physical processes at play using observationally
and/or theoretically motivated prescriptions coupled to dark matter merger
trees that can be constructed analytically or extracted from large cosmological
N-body simulations. Given our poor understanding of the physical processes
involved, and the existence of a complex interrelation between them, none of the 
SAMs that have been published matches all the statistical properties
observed \citep{neistein2010, wang2012}. In this work, in particular, we will
take advantage of two different models, with different problems. The SAM
of \citet[][DLB07]{DLB07} over-predicts the abundance of galaxies with
low and high stellar masses but reproduces the two-point galaxy
correlation functions in different stellar
mass bins (CFs) measured in the local Universe. The SAM 
of \citet[][Guo11]{guo11} matches the observed galaxy stellar mass
function in the local Universe except for the most massive end, but over-predicts the CFs for
galaxies less massive than $10^{10.77}M_{\odot}$.
The two SAMs of DLB07 and Guo11 that we use in this work are both based on the 
halo merger trees from the Millennium Simulation \citep{springel2005}.

We will also use the empirical HOD model of \citet[][hereafter Wang06]{wang2006}, 
which is also based on the Millennium Simulation. In this model, galaxy 
positions and velocities are assigned by following the orbits and merger histories of
substructures in the simulation, as is done in the SAMs. Following the empirical
HOD approach, rather than using detailed physical recipes to calculate the 
evolution of galaxy properties, galaxy
stellar mass is linked directly with the galaxy parent dark matter halo mass at the 
time of infall, assuming a double power-law function. The parameters describing the
$M_{\rm star}$ -- $M_{\rm infall}$ relation are constrained by fitting both the
SMF and the CFs from SDSS
measurements. Therefore, by construction, Wang06 can fit both the observed
SMF and the measured CFs.  

In this paper, we start by studying the $M_{\rm star}$ -- $M_{\rm infall}$
relation in the two SAMs of DLB07 and Guo11, and compare the relation with that
of Wang06. We then construct two `rescaled SAMs' based on DLB07 and Guo11, by
simply rescaling the stellar masses in SAMs so that the median and the amount of 
scatter of the $M_{\rm star}$ -- $M_{\rm infall}$ relation are the same as in
Wang06, while retaining the relative deviations of the model galaxies from the
median relation. In this way, the rescaled SAMs and Wang06 differ only on how
galaxies populate the scatter of the $M_{\rm star}$ -- $M_{\rm infall}$
relation. We demonstrate that this detail affects significantly the clustering
properties of galaxies.

This paper is organized as follows: in Sec. 2, we briefly introduce the 
models analysed in this work. In Sec. 3.1, we compare the SMF and CFs from
Wang06 with predictions from the two SAMs, and analyse the $M_{\rm star}$ --
$M_{\rm infall}$ relations of these three models.  In Sec. 3.2, we discuss the
rescaled SAMs and their predictions. In Sec. 4, we analyse the dependence on the
halo formation time of the scatter in the $M_{\rm star}$ -- $M_{\rm infall}$
relation. A discussion of our findings and our conclusions are given in
Sec. 5.

%%%%%%%%%%%%%%%%%%%%%%%%%%%%%%%%%%%%%%%%%
\section{The models}

\subsection{The Wang06 model}

As explained above, the empirical HOD model of \citet{wang2006} matches, by
construction, both the galaxy SMF and the
CFs measured in the local Universe. This model assumes a
double power law relation between the galaxy stellar mass ($M_{\rm star}$) and
the halo mass at the time of infall ($M_{\rm infall}$):
\begin{displaymath}
{{M}_{\rm star}} = \frac{2}{(\frac{{M}_{\rm infall}}{{{M}_{0}}})^{-\alpha}+(\frac{{ M}_{\rm infall}}{{{M}_{0}}})^{-\beta}}{\times}{k},
\end{displaymath}
There are five free parameters describing the relation. Besides $M_0$, $\alpha$, 
$\beta$ and $k$ shown above, at any
given value of $M_{\rm infall}$, the scatter in $\log(M_{\rm star})$ is
described assuming a Gaussian distribution with standard deviation
$\sigma$. For this work,
we have recomputed the best fit parameters matching the latest SDSS DR7 data
for the SMF and CFs (the Wang06 model was based on DR2 data). In particular, we
require our model to match 29 points in SMF, with $\log(M_{\rm
star}/h^{-2}M_{\odot})$ in the range [9.,11.9], and 119 points in CFs
measured for five stellar mass bins\footnote{27 points for bins of $\log(M_{\rm
star}/M_{\odot})< $11.27 with $r_p$ in the range [0.02, 9]$h^{-1}Mpc$, and 11 points for
the [11.27, 11.77] bin with $r_p$ in the range [0.8, 9]$h^{-1}Mpc$}. To account
for the systematic errors in the stellar mass estimates \citep{li2009},
relative errors are set to be no smaller than the relative error value at
$\log(M_{\rm star}/h^{-2}M_{\odot})$=11.35. The best-fit parameters are:
$M_0=3.43\times10^{11}h^{-1}M_{\odot}$, $\alpha=0.34$, $\beta=2.56$,
$\log{k}=10.23$ and $\sigma=0.169$ for central galaxies;
$M_0=5.23\times10^{11}h^{-1}M_{\odot}$, $\alpha=0.298$, $\beta=1.99$,
$\log{k}=10.30$ and $\sigma=0.192$ for satellite galaxies.

\subsection{The DLB07 and Guo11 model}
For details of the two SAMs analysed in this paper, we refer to the
original papers of \citet{DLB07} and \citet{guo11}. The basic ingredients of 
the two models are quite similar.
The Guo11 model differs from the DLB07 model in that it features a different treatment of 
satellite evolution and for a more efficient supernova feedback.
As mentioned above, the DLB07 and Guo11 models use the same halo merger trees
as in Wang06. In particular, the dynamical properties of galaxies and galaxy
positions are identical in DLB07 and Wang06. In Guo11, the treatment of
satellite galaxies, in particular regarding dynamical fraction and disruption, 
is slightly different, which results in slight differences in the total number 
and positions of satellite galaxies (see below). 

All model results shown below are based on dark matter halo merger trees
extracted from the Millennium Simulation \citep{springel2005} that adopts
a WMAP1 cosmology. The resolution of the
simulation corresponds to a subhalo mass limit of $\sim
10^{11}h^{-1}M_{\odot}$. As shown in \citet{guo11}, comparing SAM predictions
based on the Millennium Simulation to those based on the higher resolution
Millennium-II Simulation \citep{boylan2009}, model results converge at stellar
masses of about $6\times 10^9 M_{\odot}$. 

%%%%%%%%%%%%%%%%%%%%%%%%%%%%%%%%%%%%%%%%%%%%%%%%%%%%
\section{The relation between galaxy stellar mass and halo mass}
\subsection{Original models}
\label{sec:modelrelation}

\begin{figure*}
\bc
\hspace{-1.4cm}
\resizebox{16cm}{!}{\includegraphics{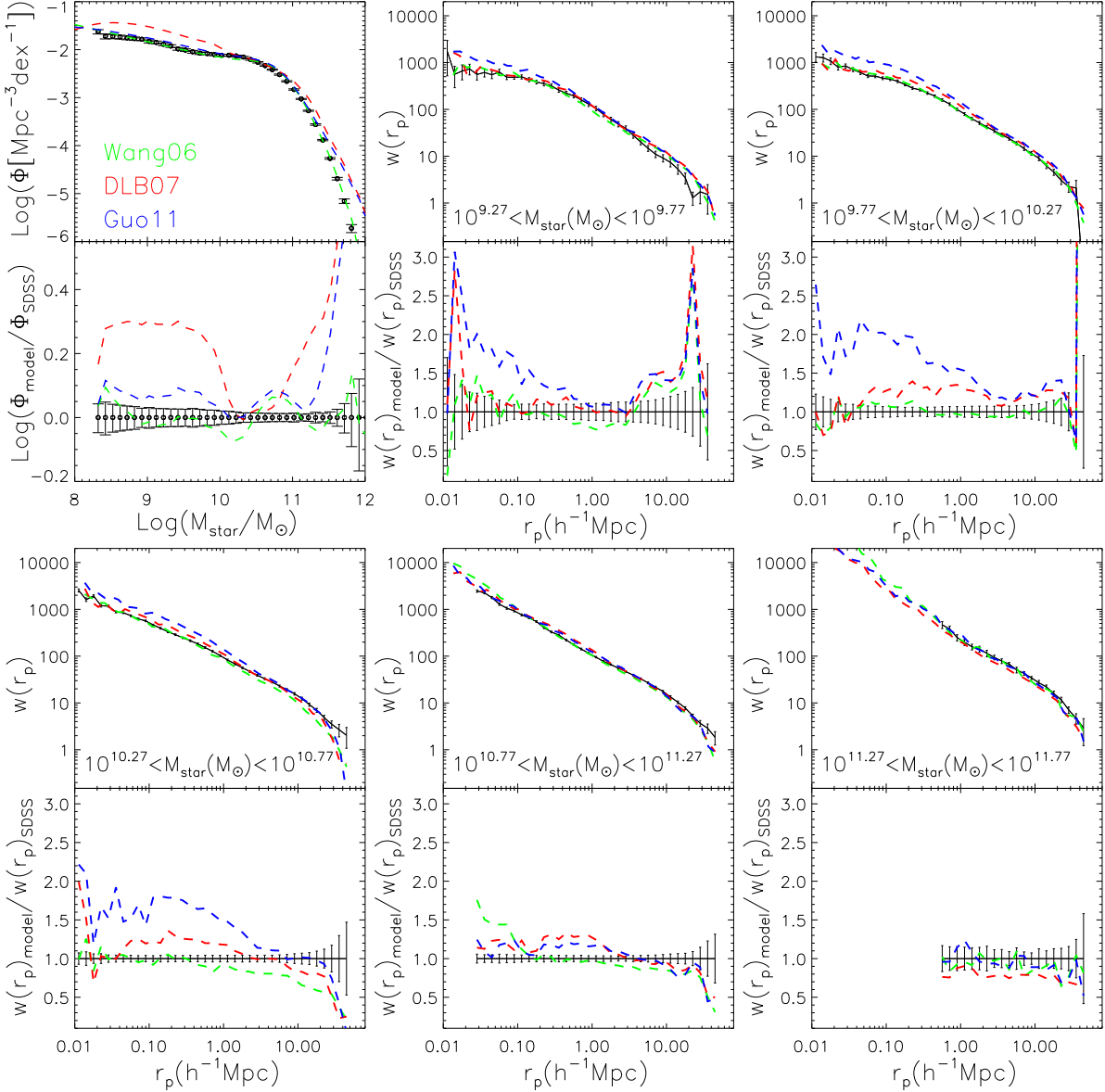}}\\%
\caption{
SMF and CFs in five stellar mass bins 
in the three models studied: the empirical HOD model of Wang06 (green lines), 
the DLB07 semi-analytic model (red lines), and the Guo11 semi-analytic model 
(blue lines). Observational results from SDSS DR7 are indicated by black 
symbols with error bars \citep{li2006,li2009,guo2010, guo11}. 
In each panel, the upper part shows the results and the lower
part gives the ratios between the models and the SDSS observation. 
By construction, the Wang06 model can reproduce both SMF and CFs, while two SAMs can not. 
}
\label{fig:modelsSMFCF}
\ec
\end{figure*}

As mentioned in Sec. 1, the Wang06 model reproduces both the SMF and CFs, while 
DLB07 and Guo11 do not fit both observations, although all the models are built 
on almost exactly the same dark matter halo merger trees. Therefore the different 
predictions for the SMF and CFs in three models considered must be due to a 
different relation between $M_{\rm star}$ and $M_{\rm infall}$. As a first step, 
in Fig.~\ref{fig:modelsSMFCF}, we show the SMF and CFs in the three original 
models, and compare them with the SDSS DR7 results \citep{li2006,li2009,guo2010, guo11}.

\begin{figure*}
\bc
\hspace{-1.4cm}
%\resizebox{15cm}{!}{\includegraphics{./MMinfallnew.ps}}\\%
\resizebox{15cm}{!}{\includegraphics{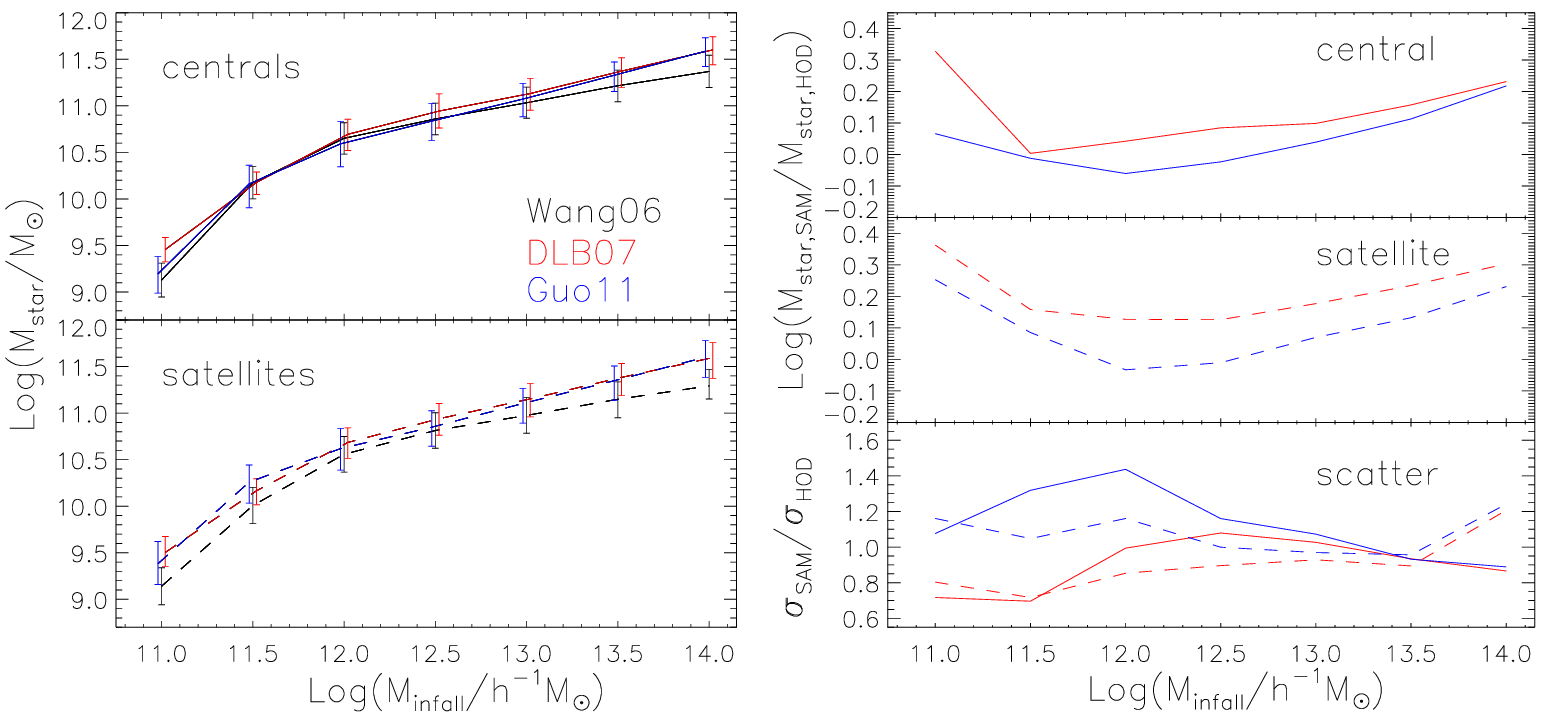}}\\%
\caption{
%Left panel: The $M_{\rm star}$ -- $M_{\rm infall}$ relations in the Wang06 HOD model, 
%DLB07 and the Guo11 semi-analytic models. The results of the DLB07 (Guo11) 
%model have been shifted down (up) by 1 dex, and $M_{\rm infall}$ of satellite 
%galaxies has been shifted by 0.03 dex, for visualization purposes. Error bars 
%show the 68 percentile distribution limits. 
Left panel: The $M_{\rm star}$ -- $M_{\rm infall}$ relations  in the Wang06 HOD model, 
DLB07 and the Guo11 semi-analytic models, for central and satellite galaxies separately.
Error bars show the 68 percentile distribution limits. 
Right panel: The ratio between the stellar mass of DLB07 (red lines)/ Guo11 
(blue lines) and that of the Wang06 HOD model as a function of $M_{\rm infall}$. 
The top and middle right panels give results of the median value for central and 
satellite galaxies. The bottom right panel indicates the ratio between the scatter 
$\sigma$ in the SAM and in the HOD model, with results for the DLB07 model shown in 
red and results for the Guo11 model in blue. Solid lines are for central galaxies, 
and dashed lines are for satellite galaxies. The semi-analytical models 
consistently predict higher stellar masses for a given halo mass, and more scatter, 
than the HOD model.}
\label{fig:mminfall}
\ec
\end{figure*}

Note that the DLB07 model was mainly constrained by the observed
K-band luminosity function, and was not tuned to reproduce the 
measured CFs. Fig.~\ref{fig:modelsSMFCF} shows that the DLB07 model actually
reproduces the measured CFs in all stellar mass bins, but over-predicts the low and high
mass end of the SMF. The Guo11 model, on the other hand, was tuned to 
reproduce the observed SMF, and therefore matches very well these observations, down to the 
lowest galaxy stellar masses measured. However, it over-predicts the number
density of the most massive galaxies, and the CFs of galaxies
less massive than $\sim 10^{10.5}h^{-2}M_{\odot}$.

We show the relation between $M_{\rm star}$ and $M_{\rm infall}$ in the 
left panel of Fig.~\ref{fig:mminfall}, for central (solid lines)
and satellite (dashed lines) galaxies. The two upper right panels show the ratio of
the median stellar mass from the two SAMs and Wang06 model as a function of
halo mass. 
Clearly, there are differences in the relation between $M_{\rm star}$ and $M_{\rm infall}$
in the three models:
(i) At fixed halo mass, satellite galaxies are less massive than centrals in the 
Wang06 model, while satellites are equally massive as centrals in the DLB07 model
and more massive than centrals at $M_{\rm infall} < 10^{12}h^{-1}M_{\odot}$ in the
Guo11 model. 
(ii) At low halo masses, both centrals and satellites in the DLB07 model are 
significantly more massive than in the Wang06 model, which results
in an excess of low-mass galaxies with respect to the observed SMF.
In the Guo11 model, at low halo masses, with a similar mass of centrals as in the 
Wang06 model, the low-mass end of the observed SMF is reproduced.
(iii) At large halo masses, both SAMs predict more massive centrals and satellites
than in the Wang06 model, and translates into an excess of massive galaxies with 
respect to the observed SMF.

\begin{figure}
\bc
\hspace{-0.8cm}
\resizebox{8.5cm}{!}{\includegraphics{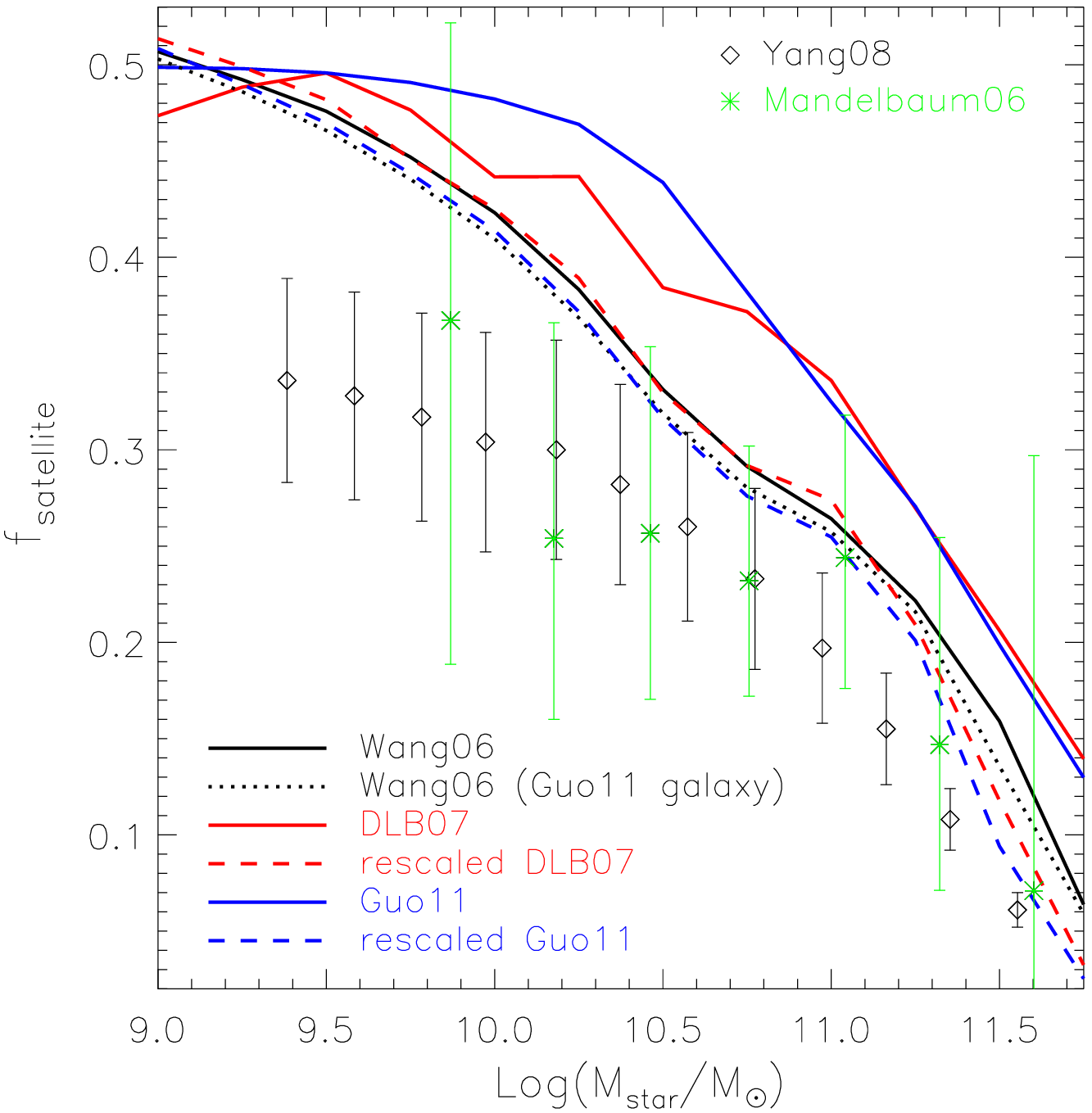}}\\%
\caption{
Satellite fractions as a function of stellar mass in different models. 
The solid black, red and blue lines show the results of the Wang06, DLB07, 
and Guo11 models. The dotted black line is the Wang06 model
combined with the galaxy information of the Guo11 model. The dashed red and 
blue lines are results of the rescaled DLB07 and Guo11 models discussed in
Section 3.2. Results
from the Yang et al. (2008) group catalogue are shown by black diamonds with 
error bars. Green stars with error bars are weak lensing results by 
\citet{mandelbaum2006}.
The Wang06 model and two rescaled SAMs give satellite fractions that 
are closer to observational measurements. 
}
\label{fig:satfraction}
\ec
\end{figure}

In the bottom right panel of Fig.~\ref{fig:mminfall}, we show the ratio of the scatter in the
$M_{\rm star}$ -- $M_{\rm infall}$ relation in the two SAMs considered to that
in HOD model. In the Wang06 HOD model, the scatter around the median $M_{\rm
star}$ -- $M_{\rm infall}$ relation is assumed to be independent of halo mass.
In the SAMs, both DLB07 and Guo11 predict larger scatter than Wang06, 
by up to $\sim 40$ per cent.

Different $M_{\rm star}$ -- $M_{\rm infall}$ relations also result in
different satellite fractions, as shown in Fig.~\ref{fig:satfraction}. 
Both the DLB07 and Guo11 models have a
higher satellite fraction than the HOD model, and the Guo11 model has a
higher fraction of satellites than DLB07 in the mass range $\log (M_{\rm
star}/M_{\odot}) \sim $ [9.5,10.8]. The differences in the satellite fractions can be
again explained by the respective $M_{\rm star}$ -- $M_{\rm infall}$ relations of
central and satellite galaxies. In DLB07, satellites are more massive than in
the HOD for any value of $M_{\rm infall}$, which results in relatively more high-mass 
satellites. In the Guo11 model, although centrals have similar
masses as in the HOD, satellites are more massive than centrals at the low-mass end,
resulting in a higher fraction of satellites. In
Fig.~\ref{fig:satfraction}, we also over-plot the measured satellite fraction
from the group catalogue of \citet{yang2008} and results from galaxy-galaxy
lensing of \citet{mandelbaum2006}. Observational uncertainties are still rather
large, but in general the HOD fractions are closest to observational results while
both SAMs predict larger satellite fractions than seen in observations 
\citep[see also][]{lu2012}. 

As noted earlier, the DLB07 and Guo11 models have slightly different satellite 
galaxy numbers/positions due to a different treatment for satellite mergers and
disruption. The HOD model presented here is based on the same dynamical
information used in the DLB07 model. We have tested that these differences do
not affect model predictions significantly: the dotted line in
Fig.~\ref{fig:satfraction} shows results obtained by repeating our fitting
procedure using the dynamical information extracted from the Guo11 model. In
this case, the satellite fraction measured in the HOD model is only about 0.01
lower than in the HOD model based on the DLB07 galaxies. This difference
is much smaller than the measured differences between SAMs and the HOD, and
between the two different SAMs.

As we have shown above, predictions from the DLB07 model are in quite good
agreement with the observed CFs, for the entire mass range
sampled by the SDSS data, despite a larger fraction of satellites
than observed. For the
Guo11 model, the predicted CFs is higher than observational
data for low-mass galaxies. Note that the three models presented in this
work are all based on the Millennium Simulation, which uses cosmological
parameters consistent with the WMAP first year result, with $\sigma_8$=$0.9$, 
higher than the latest WMAP-7 result \citep{komatsu2011}.
Guo et al. (2011) argued that the large clustering signal in their model could be 
due to the out-dated cosmology used. However, different studies have been 
carried out to investigate this issue, and these have shown that 
a lower value of $\sigma_8$ is not sufficient to bring model results 
in agreement with observational measurements \citep{wang2008,kang2012,guo2012}. 

%%%%%%%%%%%%%%%%%%%%%%%%%%%%%%%%%%%%%%%%%
\subsection{Rescaled SAMs}
\label{sec:rescaled}

In this section, we test if SAM predictions can be brought into agreement with
observational data, for both the SMF and CFs, by simply rescaling the $M_{\rm star}$ --
$M_{\rm infall}$ relation in the SAMs to be the same as in HOD model. For each
$M_{\rm infall}$ bin, we rescale the stellar masses of galaxies in SAMs to
have the same median stellar mass value, as well as the same
scatter around the median value as in the HOD. The relative deviations from the median
relation are not altered, i.e., in each halo mass bin, galaxies that are more
massive than predicted by the median relation are still more massive in the
rescaled catalogue. Satellite and central galaxies are rescaled separately. In
other words, our working assumption is that the two SAMs populate the scatter 
in the $M_{\rm star}$ -- $M_{\rm infall}$ relation correctly,
but that the absolute
value predicted for the galaxy stellar mass is offset with respect to the
correct value by an amount that is equal to the offset with respect to the HOD
median relation. 

\begin{figure*}
\bc
\hspace{-1.4cm}
\resizebox{16cm}{!}{\includegraphics{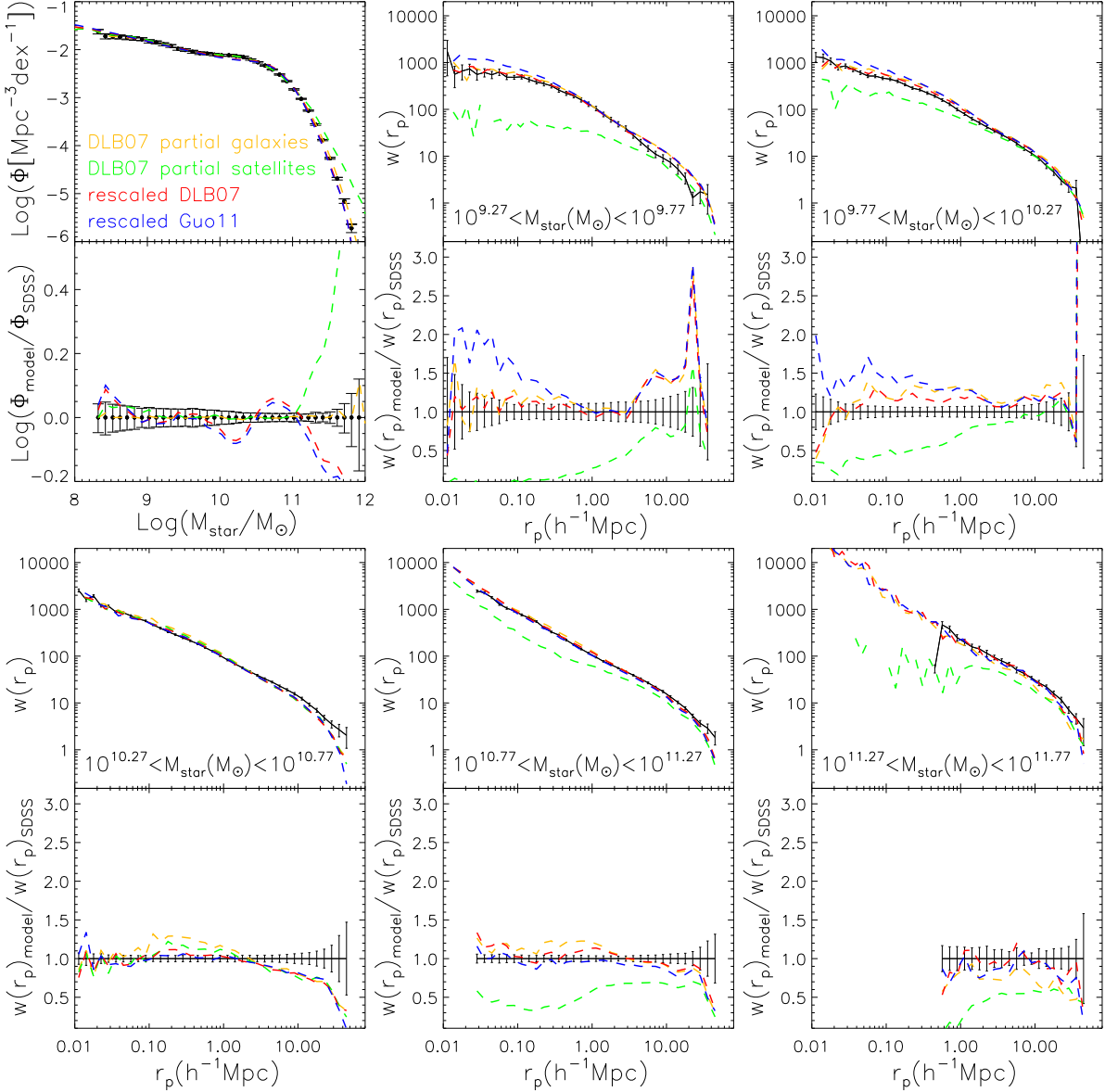}}\\%
\caption{
SMF (upper left panel) and CFs in five
stellar mass bins in different models: The DLB07 model when removing randomly a 
fraction of both centrals and satellites
(orange lines, see text for detail), DLB07 model with satellite 
galaxies partially removed (green lines, see text for detail), the rescaled 
DLB07 model (red lines), and the rescaled Guo11 model (blue lines). Note that for the
two rescaled models, both the median $M_{\rm star}$ -- $M_{\rm infall}$ relation and the 
scatter around the median are rescaled. Black symbols with error bars are 
SDSS DR7 results. The lower part of each panel shows the ratio between 
model results and observations. 
Only when part of both central and satellite galaxies are removed, DLB07
model can reproduce both SMF and CFs. Rescaling works for DLB07, but not for Guo11.
}
\label{fig:rescaledmodels}
\ec
\end{figure*}

Results of our exercise are shown in Fig.~\ref{fig:rescaledmodels}. Red lines
show the results for the rescaled DLB07 model that appears to reproduce
both the observed SMF and the CFs very well. Note that for CFs, the rescaled 
model is close to the original one, with only small differences for 
low-mass galaxies. 
When the $M_{\rm star}$ -- $M_{\rm infall}$ relation is rescaled, as
expected, the satellite fraction predicted by the rescaled model is consistent
with that of the HOD, as shown in Fig.~\ref{fig:satfraction}.

We also test two other simple models using the DLB07 predictions: in one case,
we remove randomly a fraction of galaxies in each stellar mass bin so as to
reproduce the observed galaxy SMF. Results of this exercise are shown as orange
lines in Fig.~\ref{fig:rescaledmodels}. Since the original model reproduces
quite well the observed CFs, removing randomly a subset of galaxies in each
stellar mass bin does not alter this agreement. In the other case, we remove
only satellite galaxies. Results for this case are shown in green and show
that, while the SMF is adjusted to fit observation, the CFs at small scales are 
largely suppressed. These simple tests demonstrate that, at least for the DLB07 
model, reducing the number of satellites is not the right solution to get an 
improved model that matches both the SMF and CFs. Satellite galaxies are not the 
only galaxy type to be over-abundant: the number of centrals at low-mass end 
also appears to be over-predicted in this model. Note that this over-abundance 
does not apply to galaxies in the mass range 
$\log (M_{\rm star}/M_{\odot})$ =[10.27,10.77], where the original DLB07 model 
fits both the SMF and CFs well, and only a few satellites need to be removed. 
For more massive galaxies, while the high mass end of the observed SMF is very 
uncertain \citep{bernardi2010}, the original DLB07 model can be considered 
already doing a good job in reproducing both the observed SMF and the measured CFs. 

In summary, Fig.~\ref{fig:rescaledmodels} shows that there are two possible
ways to bring the predicted SMF and CFs from the DLB07 model in agreement with
data: (1) rescale the $M_{\rm star}$ -- $M_{\rm infall}$ relation, to assign a
lower galaxy mass to low-mass haloes; (2) reduce the number of low-mass
galaxies randomly (both centrals and satellites).

The same rescaling does not work for the Guo11 model, as shown by the blue
lines in Fig.~\ref{fig:rescaledmodels}. With the same $M_{\rm star}$ -- $M_{\rm
infall}$ relation, and hence similar satellite fraction as in the HOD (dashed
blue line in Fig.~\ref{fig:satfraction}), the rescaled Guo11 model still
over-predicts the CFs at low masses. This suggests that \emph{the distributions of
galaxies within the scatter around the median $M_{\rm star}$ -- $M_{\rm
infall}$ relation affects significantly the predicted CFs.}

%%%%%%%%%%%%%%%%%%%%%%%%%%%%%%%%%%%%%%%%%
\section{The scatter of the $M_{\rm star}$ -- $M_{\rm infall}$ relation:
dependence on halo formation time} 
\label{sec:scatter}

In this section, we investigate the scatter in the $M_{\rm star}$ --
$M_{\rm infall}$ relation in detail, to understand the differences between the models
discussed in the previous section. As explained above, in the Wang06 model
galaxy stellar masses are assigned assuming a random scatter around the median
relation. In the SAMs, the scatter around the median relation is not `assumed'
but follows naturally from the scatter in the halo mass accretion history and
the stochasticity of the physical processes that drive the formation and evolution
of galaxies within haloes of fixed mass. Using predictions from the two 
SAMs, we can therefore check if and how these processes influence the
scatter in the $M_{\rm star}$ -- $M_{\rm infall}$ relation.

\subsection{Clustering for low- and high stellar mass galaxies in a fixed halo mass bin}

\begin{figure*}
\bc
\hspace{-1.4cm}
\resizebox{15cm}{!}{\includegraphics{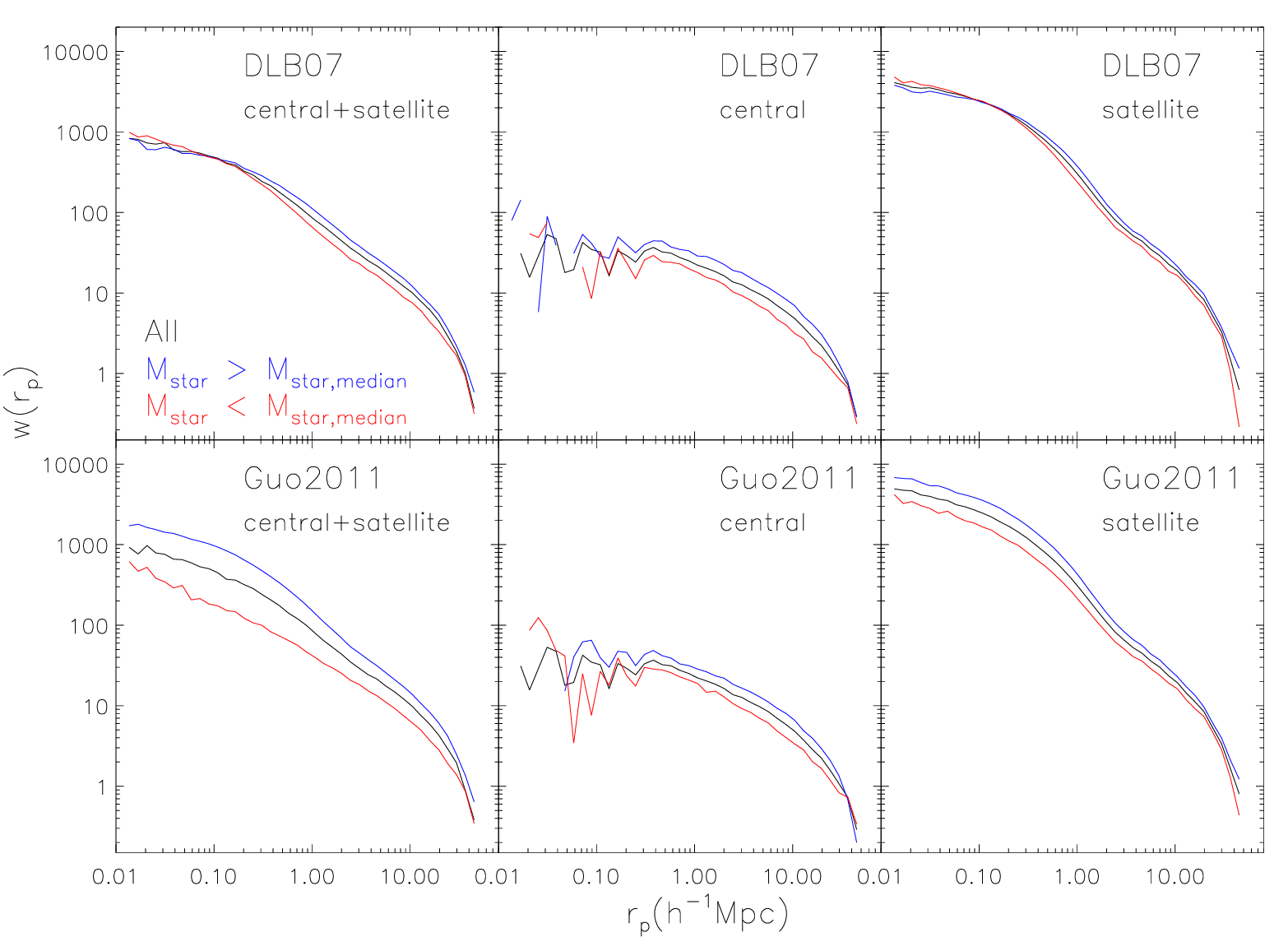}}\\%
\caption{
CFs of galaxies in subsamples split by stellar mass in the halo 
mass bin of $\log (M_{\rm infall}/h^{-1}M_{\odot})$=[11.3,11.5]. The top and bottom 
panels are for DLB07 and Guo11 models respectively. For each model, results of 
subsamples of all galaxies, central galaxies and satellite galaxies are shown from 
left to right. In each panel, the black line shows the CF for the whole sample, and 
blue/red lines show the CFs for subsamples with stellar masses above/below the median.
For both centrals and satellites, galaxies with stellar mass
above the median cluster more than the ones below the median, and the effect is stronger
in Guo11 for satellites. 
}
\label{fig:checkscatter}
\ec
\end{figure*}

As a basic check, we can simply split galaxies at a fixed halo mass into two
sub-samples according to whether the stellar mass is above or below the median 
stellar mass in the bin. In the HOD, these two sub-sample
have the same correlation function by construction (because the scatter is
random). For the SAMs, we find that this is not the case for low-mass
haloes. We show this in Fig.~\ref{fig:checkscatter} where we plot the CFs of
central and satellite galaxies in haloes with $\log (M_{\rm
infall}/h^{-1}M_{\odot})$=[11.3, 11.5]. Blue and red lines show the CFs of galaxies with stellar
mass larger and smaller than the median stellar mass of all galaxies in the
halo mass bin considered. Top panels are for the DLB07 model, while bottom
panels correspond to the Guo11 predictions. 

Fig.~\ref{fig:checkscatter} shows that, in both
models, galaxies that are more massive than the median cluster more
strongly. The difference in the clustering signal is comparable in the
two models when considering central galaxies only. For satellite galaxies, the
effect is more prominent in the Guo11 model than in the DLB07 model, and
the differences visible in the right panels of Fig.~\ref{fig:checkscatter} strongly 
influence the clustering signal for all galaxies.
We have checked that similar results are found in both SAMs combined with 
the higher resolution Millennium-II Simulation \citep{boylan2009}.
Besides, we have checked that in the same halo mass bin considered, the CFs of galaxies with
halo infall mass larger and smaller than the median differ very little.
These results show that galaxy stellar masses are not
randomly distributed within the scatter for a given halo mass \citep[see also ][]{neistein2011b}. 
The details of the scatter matter and significantly affect the
predicted CFs. We stress that the two models considered use the same dark matter 
merger trees as basic input, and mainly differ in their treatment of the 
supernovae feedback process. Therefore, our results demonstrate that the distribution 
of galaxy stellar masses with respect to the median relation can be affected 
significantly by a different modelling of baryonic physics.

\begin{figure*}
\bc
\hspace{-1.4cm}
\resizebox{16cm}{!}{\includegraphics{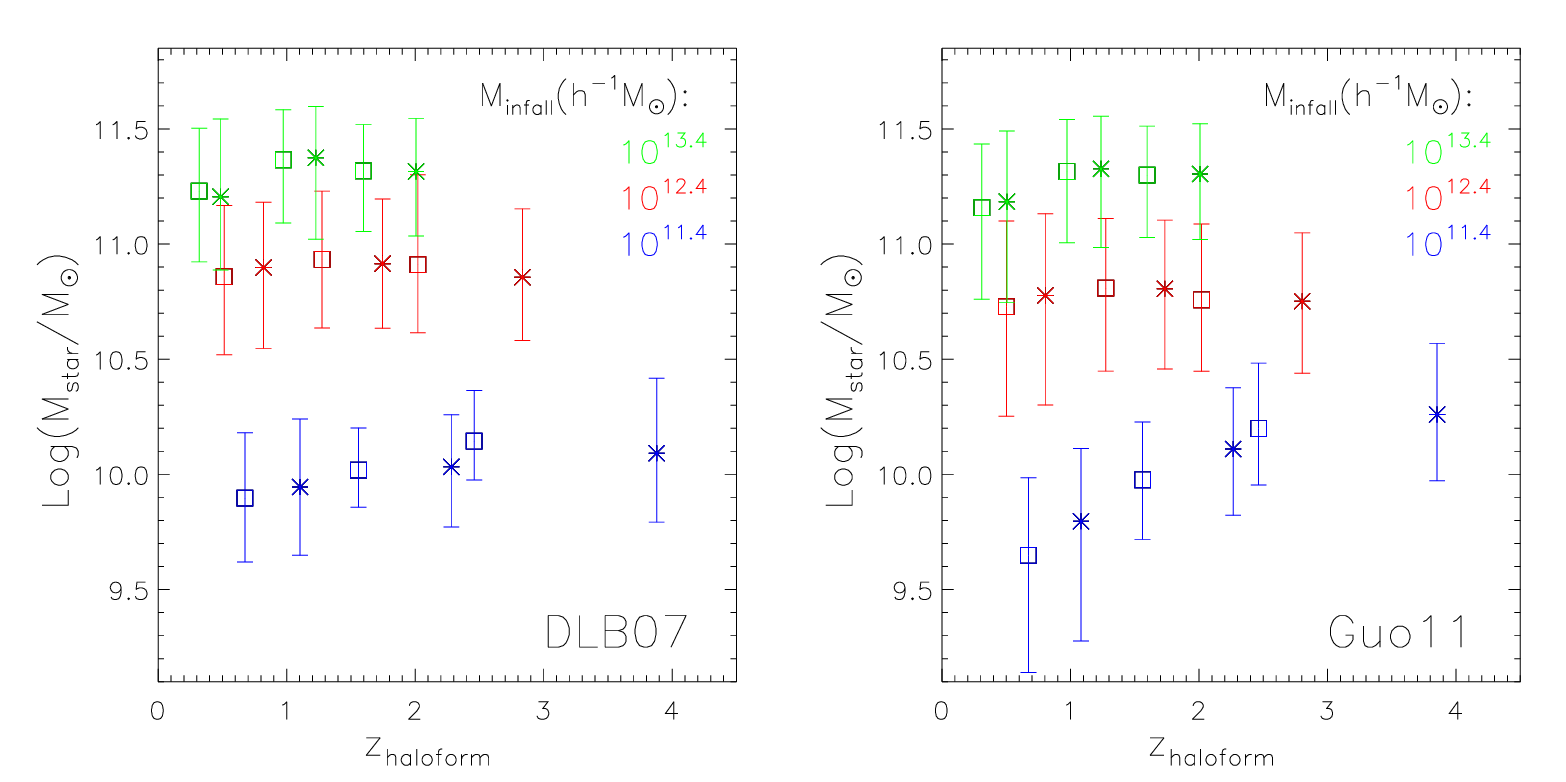}}\\%
\caption{
The relation between stellar mass and the redshift of halo formation for centrals 
(squares) and satellites (stars) in the DLB07 model and the Guo11 model, for three 
halo mass bins of $\log (M_{\rm infall}/h^{-1}M_{\odot})$=[11.3, 11.5], [12.3, 12.5]
and [13.3, 13.5]. For a given $M_{\rm infall}$, galaxies are binned according to the 
formation time of their host haloes from left to right: The 16 \% that formed latest, 
the 10 \% with formation time around the median, and the 16 \% that formed earliest.
For the lowest halo mass bin considered, there is a clear dependence of galaxy stellar mass
on halo formation time, which is stronger in Guo11 than in DLB07. 
}
\label{fig:tform}
\ec
\end{figure*}

\subsection{The influence of assembly bias}
What causes the different clustering amplitudes shown in Fig.~\ref{fig:checkscatter}? 
In Section~\ref{sec:intro}, we have discussed the existence of an assembly bias, 
which causes, at a fixed halo mass, a higher clustering amplitude for haloes that 
assembled at higher redshift. 
It seems reasonable to assume that the results shown in Fig.~\ref{fig:checkscatter} 
are related to assembly bias. We illustrate that this is indeed the case in 
Fig.~\ref{fig:tform}, where we show the relation between the stellar mass
and the halo formation time in three halo mass bins in the DLB07 and the Guo11 models. 
The halo formation time is defined as the time when $50$ per cent of the final halo 
mass is assembled in a single object.

Results in blue show the halo mass bin $\log (M_{\rm infall}/h^{-1}M_{\odot})$=[11.3, 11.5], 
which is the same mass range used in Fig.~\ref{fig:checkscatter}. We can see two clear 
trends: (i) At fixed $M_{\rm infall}$, earlier forming haloes contain more massive galaxies, 
indicating that assembly bias can indeed explain the results in 
Fig.~\ref{fig:checkscatter} and (ii) at fixed $M_{\rm infall}$, satellite galaxies on 
average form earlier \citep[see also ][]{neistein2011a}. The result is clearly more 
pronounced in Guo11 than in DLB07, indicating that the details in the baryonic physics have a 
substantial influence on the strength of assembly bias, that reflects into a dependence of
galaxy stellar mass on halo formation time. 
We have checked that the relations between stellar mass and halo formation
redshift still hold with similar slopes when using narrower halo mass bins.

We also check the same relation in two higher halo mass bins,  $\log (M_{\rm
infall}/h^{-1}M_{\odot})$=[12.3, 12.5] and [13.3, 13.5], shown in red and green respectively. 
While the result that satellites form earlier persists, we do not
anymore see a clear correlation between stellar mass and time of assembly.

In summary, we conclude that at a fixed, low halo mass, galaxies with different 
stellar masses are clustered differently: lower mass galaxies are clustered less 
than higher mass galaxies. This is because the ``over-massive galaxies'' reside 
in haloes that form early, while the ``under-massive'' galaxies are in haloes that
form late. Therefore, SAM galaxies do not populate the scatter of the 
$M_{\rm star}$ -- $M_{\rm infall}$ relation randomly. The clustering properties 
of galaxies are influenced by halo assembly bias, which is by construction not included 
in HOD models..

%%%%%%%%%%%%%%%%%%%%%%%%%%%%%%%%%%%%%%%%
\section{Discussion and Conclusions}
\label{sec:conclusion}

In this paper, we compare results from the empirical HOD model
of \citet{wang2006} with predictions from the semi-analytic models presented in 
\citet{DLB07} and \citet{guo11}, both based on the halo merger trees extracted
from the Millennium Simulation. By construction, the HOD model is able to
reproduce simultaneously the galaxy SMF and the 
CFs, down to the stellar mass limit of the SDSS. The semi-analytic
models have problems in reproducing both these observations. In particular, the
DLB07 model reproduces quite well the dependence of the clustering amplitude on
mass but over-predicts the number densities of low-to-intermediate mass
galaxies. In contrast, the Guo11 model reproduces the stellar galaxy
mass function down to the lowest mass measured (it does so by construction),
but over-predicts the clustering amplitude for low-mass galaxies. These
different predictions can be explained by comparing the predicted $M_{\rm
star}$ -- $M_{\rm infall}$ relations with that obtained by the HOD approach. 

We demonstrate that scaling the results from the semi-analytic model so as to
force them to reproduce the same $M_{\rm star}$ -- $M_{\rm infall}$ relation
that is found in the HOD does not suffice to bring them in agreement with both
observational measurements used to constrain the HOD. Instead, we show that the way model
galaxies populate the scatter around the median relation matters. In the HOD
model, as in most other models that are found in the literature, the
scatter around the $M_{\rm star}$ -- $M_{\rm infall}$ relation is modelled as a
random Gaussian distribution. In the semi-analytic models we use, stellar masses exhibit
clear dependence on halo formation time, with stronger trends for low-mass
galaxies. At given $M_{\rm infall}$, galaxies with larger stellar mass reside
in haloes that formed earlier and consequently have a higher clustering amplitude than
haloes with the same mass but later formation times \citep{gao2005}. The
influence of assembly bias on galaxies is stronger in the Guo11 
model than in the DLB07 model, and
results in an excess of the clustering signal for low-mass galaxies. 

Does assembly bias exist in the real Universe? The issue is still matter of
debate. \citet{tinker2008} conclude there is no evidence for assembly bias for 
low-mass galaxies from the fact that HOD models match the observed void statistics 
of red and blue galaxies. \citet{skibba2009} studied the environmental 
dependence of galaxy colours, and argued that the effects of 
assembly bias are probably small. 
If the effect on galaxies is present at the levels found 
in the DLB07 model for central and especially satellite galaxies, 
it might be difficult to distinguish it from just a random scatter using 
observational constraints as the SMF and CFs in different stellar mass bins.
Measurements of correlation function for galaxies in fixed stellar mass bins
but split by colour and/or specific star formation rate may help to answer 
this question. We address this issue in a companion paper.

If assembly bias significantly affects galaxies in the real Universe, as it 
does in the SAMs, one needs to be very careful when applying models 
neglecting this effect, like HOD and abundance matching models, in particular
regarding low-mass galaxies and satellites \citep{zu2008}. For example, for a given 
$M_{\rm star}$ -- $M_{\rm infall}$ relation and a given scatter around that 
relation, assuming random scatter will produce lower CFs than 
assuming a scatter accounting for assembly bias. If the correlation
function is then reproduced by coincidence one may draw wrong conclusions about
the importance of other effects that should have made clustering less strong,
such as tidal stripping and reduced merger times of galaxies. 
Finally, if significant, assembly bias is relevant for precision measurements 
of cosmological parameters \citep[e.g.][]{zu2008}. Future HOD and abundance
matching models would need to account for a non-random scatter including the
assembly bias effect. 

%%%%%%%%%%%%%%%%%%%%%%%%%%%%%%%%%%%%%%%%
\section*{Acknowledgments}

LW acknowledges support from the National basic research program
of China (973 program under grant No. 2009CB24901), the NSFC grants program 
(No. 11143006, No. 11103033, No. 11133003),
and the Partner Group program of the Max Planck Society.
GDL acknowledges financial support from the European Research Council under the
European Community's Seventh Framework Programme (FP7/2007-2013)/ERC grant
agreement n. 202781. 
SMW acknowledges funding from ERC grant HIGHZ no. 227749.

The simulation used in this paper was carried out as part of the programme of
the Virgo Consortium on the Regatta supercomputer of the Computing Centre of
the Max–Planck–Society in Garching.  The halo data, together with the galaxy
data from two semi-analytic galaxy formation models, are publicly available at
http://www.mpa-garching.mpg.de/millennium/.

\bsp
\label{lastpage}

\bibliographystyle{mn2e}
\bibliography{scatter}

\begin{thebibliography}{}

\bibitem[\protect\citeauthoryear{{Benson}, {Cole}, {Frenk}, {Baugh} \&
  {Lacey}}{{Benson} et~al.}{2000}]{benson2000}
{Benson} A.~J.,  {Cole} S.,  {Frenk} C.~S.,  {Baugh} C.~M.,    {Lacey} C.~G.,
  2000, \mnras, 311, 793

\bibitem[\protect\citeauthoryear{{Berlind} \& {Weinberg}}{{Berlind} \&
  {Weinberg}}{2002}]{berlind2002}
{Berlind} A.~A.,  {Weinberg} D.~H.,  2002, \apj, 575, 587

\bibitem[\protect\citeauthoryear{{Berlind}, {Weinberg}, {Benson}, {Baugh},
  {Cole}, {Dav{\'e}}, {Frenk}, {Jenkins}, {Katz} \& {Lacey}}{{Berlind}
  et~al.}{2003}]{berlind2003}
{Berlind} A.~A.,  {Weinberg} D.~H.,  {Benson} A.~J.,  {Baugh} C.~M.,  {Cole}
  S.,  {Dav{\'e}} R.,  {Frenk} C.~S.,  {Jenkins} A.,  {Katz} N.,    {Lacey}
  C.~G.,  2003, \apj, 593, 1

\bibitem[\protect\citeauthoryear{{Bernardi}, {Shankar}, {Hyde}, {Mei},
  {Marulli} \& {Sheth}}{{Bernardi} et~al.}{2010}]{bernardi2010}
{Bernardi} M.,  {Shankar} F.,  {Hyde} J.~B.,  {Mei} S.,  {Marulli} F.,
  {Sheth} R.~K.,  2010, \mnras, 404, 2087

\bibitem[\protect\citeauthoryear{{Boylan-Kolchin}, {Springel}, {White},
  {Jenkins} \& {Lemson}}{{Boylan-Kolchin} et~al.}{2009}]{boylan2009}
{Boylan-Kolchin} M.,  {Springel} V.,  {White} S.~D.~M.,  {Jenkins} A.,
  {Lemson} G.,  2009, \mnras, 398, 1150

\bibitem[\protect\citeauthoryear{{Conroy}, {Wechsler} \& {Kravtsov}}{{Conroy}
  et~al.}{2006}]{conroy2005}
{Conroy} C.,  {Wechsler} R.~H.,    {Kravtsov} A.~V.,  2006, \apj, 647, 201

\bibitem[\protect\citeauthoryear{{Croton}, {Gao} \& {White}}{{Croton}
  et~al.}{2007}]{croton2007}
{Croton} D.~J.,  {Gao} L.,    {White} S.~D.~M.,  2007, \mnras, 374, 1303

\bibitem[\protect\citeauthoryear{{De Lucia} \& {Blaizot}}{{De Lucia} \&
  {Blaizot}}{2007}]{DLB07}
{De Lucia} G.,  {Blaizot} J.,  2007, \mnras, 375, 2

\bibitem[\protect\citeauthoryear{{Gao}, {Springel} \& {White}}{{Gao}
  et~al.}{2005}]{gao2005}
{Gao} L.,  {Springel} V.,    {White} S.~D.~M.,  2005, \mnras, 363, L66

\bibitem[\protect\citeauthoryear{{Goto}}{{Goto}}{2003}]{goto2003}
{Goto} T.,  2003, PhD thesis, The University of Tokyo

\bibitem[\protect\citeauthoryear{{Guo}, {White}, {Angulo}, {Henriques},
  {Lemson}, {Boylan-Kolchin}, {Thomas} \& {Short}}{{Guo}
  et~al.}{2012}]{guo2012}
{Guo} Q.,  {White} S.,  {Angulo} R.~E.,  {Henriques} B.,  {Lemson} G.,
  {Boylan-Kolchin} M.,  {Thomas} P.,    {Short} C.,  2012, ArXiv e-prints

\bibitem[\protect\citeauthoryear{{Guo}, {White}, {Boylan-Kolchin}, {De Lucia},
  {Kauffmann}, {Lemson}, {Li}, {Springel} \& {Weinmann}}{{Guo}
  et~al.}{2011}]{guo11}
{Guo} Q.,  {White} S.,  {Boylan-Kolchin} M.,  {De Lucia} G.,  {Kauffmann} G.,
  {Lemson} G.,  {Li} C.,  {Springel} V.,    {Weinmann} S.,  2011, \mnras, 413,
  101

\bibitem[\protect\citeauthoryear{{Guo}, {White}, {Li} \&
  {Boylan-Kolchin}}{{Guo} et~al.}{2010}]{guo2010}
{Guo} Q.,  {White} S.,  {Li} C.,    {Boylan-Kolchin} M.,  2010, \mnras, 404,
  1111

\bibitem[\protect\citeauthoryear{{Jaff{\'e}}, {Arag{\'o}n-Salamanca},
  {Kuntschner}, {Bamford}, {Hoyos}, {De Lucia}, {Halliday}, {Milvang-Jensen},
  {Poggianti}, {Rudnick}, {Saglia}, {Sanchez-Blazquez} \&
  {Zaritsky}}{{Jaff{\'e}} et~al.}{2011}]{jaffe2011}
{Jaff{\'e}} Y.~L.,  {Arag{\'o}n-Salamanca} A.,  {Kuntschner} H.,  {Bamford} S.,
   {Hoyos} C.,  {De Lucia} G.,  {Halliday} C.,  {Milvang-Jensen} B.,
  {Poggianti} B.,  {Rudnick} G.,  {Saglia} R.~P.,  {Sanchez-Blazquez} P.,
  {Zaritsky} D.,  2011, \mnras, 417, 1996

\bibitem[\protect\citeauthoryear{{Kang}, {Li}, {Lin} \& {Elahi}}{{Kang}
  et~al.}{2012}]{kang2012}
{Kang} X.,  {Li} M.,  {Lin} W.~P.,    {Elahi} P.~J.,  2012, \mnras, 422, 804

\bibitem[\protect\citeauthoryear{{Kauffmann}, {Heckman}, {De Lucia},
  {Brinchmann}, {Charlot}, {Tremonti}, {White} \& {Brinkmann}}{{Kauffmann}
  et~al.}{2006}]{kauffmann2006}
{Kauffmann} G.,  {Heckman} T.~M.,  {De Lucia} G.,  {Brinchmann} J.,  {Charlot}
  S.,  {Tremonti} C.,  {White} S.~D.~M.,    {Brinkmann} J.,  2006, \mnras, 367,
  1394

\bibitem[\protect\citeauthoryear{{Komatsu}, {Smith}, {Dunkley}, {Bennett},
  {Gold}, {Hinshaw}, {Jarosik}, {Larson} \& {et al.,}}{{Komatsu}
  et~al.}{2011}]{komatsu2011}
{Komatsu} E.,  {Smith} K.~M.,  {Dunkley} J.,  {Bennett} C.~L.,  {Gold} B.,
  {Hinshaw} G.,  {Jarosik} N.,  {Larson} D.,    {et al.,} 2011, \apjs, 192, 18

\bibitem[\protect\citeauthoryear{{Li}, {Kauffmann}, {Jing}, {White},
  {B{\"o}rner} \& {Cheng}}{{Li} et~al.}{2006}]{li2006}
{Li} C.,  {Kauffmann} G.,  {Jing} Y.~P.,  {White} S.~D.~M.,  {B{\"o}rner} G.,
   {Cheng} F.~Z.,  2006, \mnras, 368, 21

\bibitem[\protect\citeauthoryear{{Li} \& {White}}{{Li} \&
  {White}}{2009}]{li2009}
{Li} C.,  {White} S.~D.~M.,  2009, \mnras, 398, 2177

\bibitem[\protect\citeauthoryear{{Li}, {Mo}, {van den Bosch} \& {Lin}}{{Li}
  et~al.}{2007}]{liyun2007}
{Li} Y.,  {Mo} H.~J.,  {van den Bosch} F.~C.,    {Lin} W.~P.,  2007, \mnras,
  379, 689

\bibitem[\protect\citeauthoryear{{Lu}, {Mo}, {Katz} \& {Weinberg}}{{Lu}
  et~al.}{2012}]{lu2012}
{Lu} Y.,  {Mo} H.~J.,  {Katz} N.,    {Weinberg} M.~D.,  2012, \mnras, 421, 1779

\bibitem[\protect\citeauthoryear{{Mandelbaum}, {Seljak}, {Kauffmann}, {Hirata}
  \& {Brinkmann}}{{Mandelbaum} et~al.}{2006}]{mandelbaum2006}
{Mandelbaum} R.,  {Seljak} U.,  {Kauffmann} G.,  {Hirata} C.~M.,    {Brinkmann}
  J.,  2006, \mnras, 368, 715

\bibitem[\protect\citeauthoryear{{More}, {van den Bosch} \& {Cacciato}}{{More}
  et~al.}{2009a}]{more2009}
{More} S.,  {van den Bosch} F.~C.,    {Cacciato} M.,  2009a, \mnras, 392, 917

\bibitem[\protect\citeauthoryear{{More}, {van den Bosch}, {Cacciato}, {Mo},
  {Yang} \& {Li}}{{More} et~al.}{2009b}]{more2009b}
{More} S.,  {van den Bosch} F.~C.,  {Cacciato} M.,  {Mo} H.~J.,  {Yang} X.,
  {Li} R.,  2009b, \mnras, 392, 801

\bibitem[\protect\citeauthoryear{{More}, {van den Bosch}, {Cacciato}, {Skibba},
  {Mo} \& {Yang}}{{More} et~al.}{2011}]{more2011}
{More} S.,  {van den Bosch} F.~C.,  {Cacciato} M.,  {Skibba} R.,  {Mo} H.~J.,
   {Yang} X.,  2011, \mnras, 410, 210

\bibitem[\protect\citeauthoryear{{Moster}, {Somerville}, {Maulbetsch}, {van den
  Bosch}, {Macci{\`o}}, {Naab} \& {Oser}}{{Moster} et~al.}{2010}]{moster2010}
{Moster} B.~P.,  {Somerville} R.~S.,  {Maulbetsch} C.,  {van den Bosch} F.~C.,
  {Macci{\`o}} A.~V.,  {Naab} T.,    {Oser} L.,  2010, \apj, 710, 903

\bibitem[\protect\citeauthoryear{{Neistein}, {Li}, {Khochfar}, {Weinmann},
  {Shankar} \& {Boylan-Kolchin}}{{Neistein} et~al.}{2011a}]{neistein2011a}
{Neistein} E.,  {Li} C.,  {Khochfar} S.,  {Weinmann} S.~M.,  {Shankar} F.,
  {Boylan-Kolchin} M.,  2011a, \mnras, 416, 1486

\bibitem[\protect\citeauthoryear{{Neistein} \& {Weinmann}}{{Neistein} \&
  {Weinmann}}{2010}]{neistein2010}
{Neistein} E.,  {Weinmann} S.~M.,  2010, \mnras, 405, 2717

\bibitem[\protect\citeauthoryear{{Neistein}, {Weinmann}, {Li} \&
  {Boylan-Kolchin}}{{Neistein} et~al.}{2011b}]{neistein2011b}
{Neistein} E.,  {Weinmann} S.~M.,  {Li} C.,    {Boylan-Kolchin} M.,  2011b,
  \mnras, 414, 1405

\bibitem[\protect\citeauthoryear{{Peacock} \& {Smith}}{{Peacock} \&
  {Smith}}{2000}]{peacock2000}
{Peacock} J.~A.,  {Smith} R.~E.,  2000, \mnras, 318, 1144

\bibitem[\protect\citeauthoryear{{Seljak}}{{Seljak}}{2000}]{seljak2000}
{Seljak} U.,  2000, \mnras, 318, 203

\bibitem[\protect\citeauthoryear{{Skibba} \& {Sheth}}{{Skibba} \&
  {Sheth}}{2009}]{skibba2009}
{Skibba} R.~A.,  {Sheth} R.~K.,  2009, \mnras, 392, 1080

\bibitem[\protect\citeauthoryear{{Skibba}, {van den Bosch}, {Yang}, {More},
  {Mo} \& {Fontanot}}{{Skibba} et~al.}{2011}]{skibba2011}
{Skibba} R.~A.,  {van den Bosch} F.~C.,  {Yang} X.,  {More} S.,  {Mo} H.,
  {Fontanot} F.,  2011, \mnras, 410, 417

\bibitem[\protect\citeauthoryear{{Springel}, {White}, {Jenkins}, {Frenk},
  {Yoshida}, {Gao}, {Navarro}, {Thacker} \& {et al.,}}{{Springel}
  et~al.}{2005}]{springel2005}
{Springel} V.,  {White} S.~D.~M.,  {Jenkins} A.,  {Frenk} C.~S.,  {Yoshida} N.,
   {Gao} L.,  {Navarro} J.,  {Thacker} R.,    {et al.,} 2005, \nat, 435, 629

\bibitem[\protect\citeauthoryear{{Tanvuia}, {Zeilinger}, {Focardi}, {Kelm} \&
  {Rampazzo}}{{Tanvuia} et~al.}{2003}]{tanvuia2003}
{Tanvuia} L.,  {Zeilinger} W.~W.,  {Focardi} P.,  {Kelm} B.,    {Rampazzo} R.,
  2003, \apss, 284, 459

\bibitem[\protect\citeauthoryear{{Tinker}, {Conroy}, {Norberg}, {Patiri},
  {Weinberg} \& {Warren}}{{Tinker} et~al.}{2008}]{tinker2008}
{Tinker} J.~L.,  {Conroy} C.,  {Norberg} P.,  {Patiri} S.~G.,  {Weinberg}
  D.~H.,    {Warren} M.~S.,  2008, \apj, 686, 53

\bibitem[\protect\citeauthoryear{{Vale} \& {Ostriker}}{{Vale} \&
  {Ostriker}}{2006}]{vale2005}
{Vale} A.,  {Ostriker} J.~P.,  2006, \mnras, 371, 1173

\bibitem[\protect\citeauthoryear{{Valle}, {Shore} \& {Galli}}{{Valle}
  et~al.}{2005}]{valle2005}
{Valle} G.,  {Shore} S.~N.,    {Galli} D.,  2005, \aap, 435, 551

\bibitem[\protect\citeauthoryear{{Wang}, {De Lucia}, {Kitzbichler} \&
  {White}}{{Wang} et~al.}{2008}]{wang2008}
{Wang} J.,  {De Lucia} G.,  {Kitzbichler} M.~G.,    {White} S.~D.~M.,  2008,
  \mnras, 384, 1301

\bibitem[\protect\citeauthoryear{{Wang}, {Li}, {Kauffmann} \& {De
  Lucia}}{{Wang} et~al.}{2006}]{wang2006}
{Wang} L.,  {Li} C.,  {Kauffmann} G.,    {De Lucia} G.,  2006, \mnras, 371, 537

\bibitem[\protect\citeauthoryear{{Wang}, {Weinmann} \& {Neistein}}{{Wang}
  et~al.}{2012}]{wang2012}
{Wang} L.,  {Weinmann} S.~M.,    {Neistein} E.,  2012, \mnras, 421, 3450

\bibitem[\protect\citeauthoryear{{White} \& {Frenk}}{{White} \&
  {Frenk}}{1991}]{white1991}
{White} S.~D.~M.,  {Frenk} C.~S.,  1991, \apj, 379, 52

\bibitem[\protect\citeauthoryear{{Yang}, {Mo} \& {van den Bosch}}{{Yang}
  et~al.}{2003}]{yang2003}
{Yang} X.,  {Mo} H.~J.,    {van den Bosch} F.~C.,  2003, \mnras, 339, 1057

\bibitem[\protect\citeauthoryear{{Yang}, {Mo} \& {van den Bosch}}{{Yang}
  et~al.}{2008}]{yang2008}
{Yang} X.,  {Mo} H.~J.,    {van den Bosch} F.~C.,  2008, \apj, 676, 248

\bibitem[\protect\citeauthoryear{{Zehavi}, {Zheng}, {Weinberg}, {Frieman},
  {Berlind}, {Blanton}, {Scoccimarro}, {Sheth} \& {et al.,}}{{Zehavi}
  et~al.}{2005}]{zehavi2005}
{Zehavi} I.,  {Zheng} Z.,  {Weinberg} D.~H.,  {Frieman} J.~A.,  {Berlind}
  A.~A.,  {Blanton} M.~R.,  {Scoccimarro} R.,  {Sheth} R.~K.,    {et al.,}
  2005, \apj, 630, 1

\bibitem[\protect\citeauthoryear{{Zhao}, {Jing}, {Mo} \& {B{\"o}rner}}{{Zhao}
  et~al.}{2009}]{zhao2009}
{Zhao} D.~H.,  {Jing} Y.~P.,  {Mo} H.~J.,    {B{\"o}rner} G.,  2009, \apj, 707,
  354

\bibitem[\protect\citeauthoryear{{Zhao}, {Mo}, {Jing} \& {B{\"o}rner}}{{Zhao}
  et~al.}{2003}]{zhao2003}
{Zhao} D.~H.,  {Mo} H.~J.,  {Jing} Y.~P.,    {B{\"o}rner} G.,  2003, \mnras,
  339, 12

\bibitem[\protect\citeauthoryear{{Zhu}, {Zheng}, {Lin}, {Jing}, {Kang} \&
  {Gao}}{{Zhu} et~al.}{2006}]{zhu2006}
{Zhu} G.,  {Zheng} Z.,  {Lin} W.~P.,  {Jing} Y.~P.,  {Kang} X.,    {Gao} L.,
  2006, \apjl, 639, L5

\bibitem[\protect\citeauthoryear{{Zu}, {Zheng}, {Zhu} \& {Jing}}{{Zu}
  et~al.}{2008}]{zu2008}
{Zu} Y.,  {Zheng} Z.,  {Zhu} G.,    {Jing} Y.~P.,  2008, \apj, 686, 41

\end{thebibliography}

%%%%%%%%%%%%%%%%%%%%%%%%%%%%%%%%%%%%%%%%%%%%%%%%%%%%%%%%%%%%%%%%%%%%%%%%%%%%
\end{document}